# Cross-Cultural Communication in the Digital Age: An Analysis of Cultural Representation and Inclusivity in Emojis

Li Lingfeng, Zheng Xiangwen

June 2024

## Abstract

Emojis have become a universal language in the digital world, enabling users to express emotions, ideas, and identities across diverse cultural contexts. As emojis incorporate more cultural symbols and diverse representations, they play a crucial role in cross-cultural communication. This research project aims to analyze the representation of different cultures in emojis, investigate how emojis facilitate cross-cultural communication and promote inclusivity, and explore the impact of emojis on understanding and interpretation in different cultural contexts.

# Contents

# 1  Introduction

Emojis, originally created in Japan in the late 1990s, have become a universal language in the digital world. The term *emoji* comes from the Japanese words 絵 (*e*, meaning *picture*) and 文字 (*moji*, meaning *character*). They have evolved significantly over the years, expanding from a set of 176 icons to over 3,000, reflecting a wide range of emotions, objects, activities, and cultural symbols.

As digital communication has become increasingly global, emojis have incorporated more cultural terms coming from various parts of the world. This includes symbols representing food, festivals, landmarks, and other aspects unique to different cultures. Furthermore, emojis now offer diverse skin tones for people of different ethnicities to better represent themselves, and gender options to promote gender equality and care for the sex minority. There are also options for people with disabilities, reflecting a broader commitment to inclusivity.

However, despite these advancements, the use and interpretation of emojis can still vary significantly across different cultures. This can potentially lead to misunderstandings in cross-cultural communication. Therefore, it is crucial to understand how emojis are used and perceived in different cultural contexts, and how they can facilitate or hinder cross-cultural communication.

# 2  Literature Review

In the digital age, emojis have emerged as a significant aspect of cross-cultural communication, shaping the way individuals express emotions, ideas, and identities across diverse cultural contexts. A comprehensive literature review reveals that the study of emojis has gained momentum over the past few years, with an increasing number of publications addressing their development, usage patterns, functional attributes, and applications across various fields (Bai et al.). This review synthesizes key findings to provide insights into cultural representation and inclusivity within the emoji landscape.

Emojis originated as a response to the need for non-verbal cues in computer-mediated communication (CMC), where traditional forms of expression like tone, facial expressions, and gestures are limited (Archer and Akert; Harris and Paradice; Riordan). Researchers have found that emojis compensate for this lack, functioning as visual symbols that mirror real-life expressions and enhance emotional clarity in digital conversations (Tossell et al.; Negishi).

One crucial dimension of emoji research pertains to their emotional and semantic functions. Studies have classified emojis into positive, neutral, and negative categories, with the majority conveying positive emotions (Petra et al.). This suggests that emojis play a vital role in fostering positive sentiment in digital interactions. Moreover, the strategic combination of emojis can amplify emotional depth and complexity in messages (López and Cap), demonstrating their versatility as a tool for nuanced emotional expression.

Cultural representation in emojis is a focal point of recent research. The democratization of emoji selection and the introduction of emojis representing different skin tones and previously underrepresented experiences, such as menstruation, highlight efforts to enhance inclusivity (Sweeney and Whaley; 8. Lee et al.). This move towards greater representation acknowledges and addresses the need for emojis to reflect the diversity of users' experiences and identities. However, these developments also underscore the importance of understanding how emojis are perceived and utilized across cultures and how they might contribute to or mitigate social inequalities.

Individual differences, cultural backgrounds, and platform-specific factors significantly influence emoji use. Gender, age, and nationality are among the demographic characteristics impacting emoji preferences and frequency of use (Tossell et al.; Prada et al.; Li et al.). For instance, women are reported to use emojis more frequently in public settings, while men employ a wider range (Derks et al.). Furthermore, emojis carry specific national or ethnic meanings, indicating that their interpretation and adoption vary across cultures (Gaspar et al.).

Cross-cultural communication research has explored how emojis are integrated into various languages and communication styles, revealing that different language types influ-ence emoji use (Lin and Chen). This highlights the need for understanding the contextual sensitivity of emojis and their role in shaping digital discourse across cultures.

The literature emphasizes the importance of studying emojis from multidisciplinary perspectives, considering their impact on communication dynamics, psychological perceptions, marketing strategies, and societal attitudes. Future research should delve deeper into the complexities of emoji usage in diverse cultural contexts, examining their potential for promoting understanding and unity amidst global digital communities.

## 3   Methods and Procedures

This research project employed a mixed-methods approach to investigate the representation of different cultures in emojis, explore the role of emojis in cross-cultural communication, and analyze the impact of emojis on understanding and interpretation in diverse cultural contexts. The study involved a survey conducted among 106 participants (most of whom are college students in China, and have the experience of using emojis), focusing on their emoji usage patterns, preferences, and interpretations. The survey data was analyzed using qualitative methods to identify themes and patterns in participants' responses.

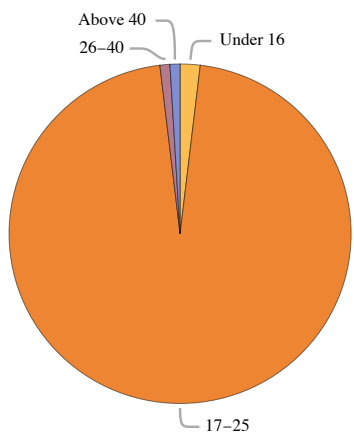

Figure 1: Age distribution of survey participants

In addition to the primary survey, this research project analyzed data from existing studies on emoji usage, cultural representation, and cross-cultural communication. By synthesizing findings from various sources, the study aimed to provide a comprehensive overview of the current state of emoji research and its implications for cross-cultural communication.

# 4 Data Analysis

In the survey conducted for this research project, participants were asked a series of questions related to understanding of emojis' cultural characteristics, the awareness of cultural differences in emoji interpretation, and the impact of emojis on cross-cultural communication. The responses were analyzed to identify common themes and patterns, shedding light on the role of emojis in facilitating cross-cultural communication and promoting cultural inclusivity.

## 4.1 Awareness of emojis with cultural characteristics

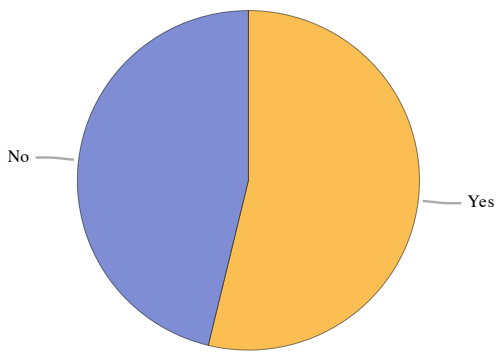

Figure 2: Awareness of multi-color options in emojis

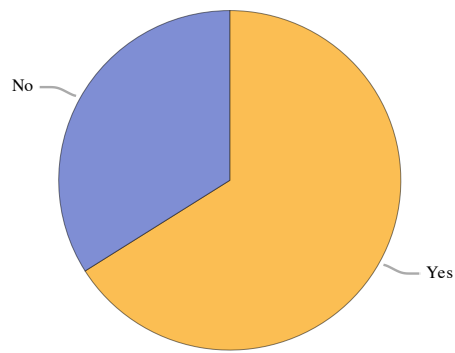

Figure 3: Noticing emojis with cultural symbols

According to the results, 53% of participants know that emojis have multi-color options, while the remaining 47% are unaware of this feature. This indicates that there is a lack of awareness among some users regarding the cultural diversity represented in emojis.

When asked about the whether they have noticed emojis with cultural symbols, 66% of participants responded affirmatively, suggesting that cultural representation in emojis is becoming more prominent and noticeable.

## 4.2   Usage of emojis in cross-cultural communication

Participants were asked about their use of emojis with exo-cultural characteristics (i.e., cultural symbols from other cultures, in this case beyond traditional Chinese culture). The results showed that more than half (69%) of the participants use them occasionally or frequently. This indicates that emojis with cultural characteristics are commonly used in daily communication, reflecting the globalization of digital communication practices.

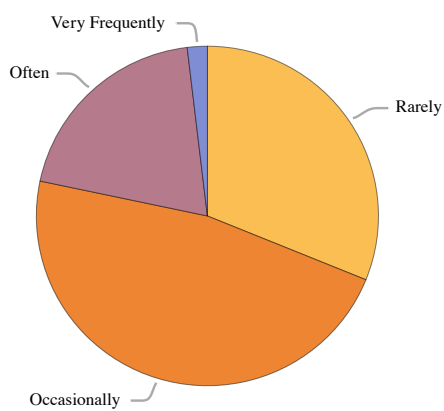
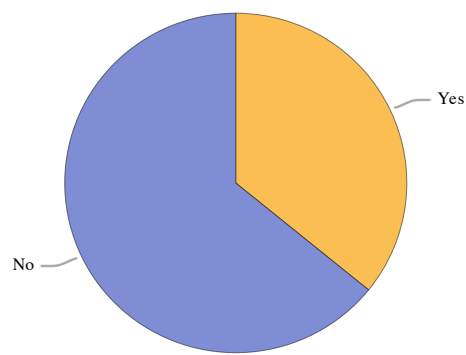

Figure 4:  Usage of emojis with exo-cultural characteristics

Figure 5: Use of emojis in cross-cultural communication

When it comes to the use of emojis in cross cultural communication, only 38% of the participants have used emojis during a cross-cultural conversation. However, as 25% of the participants have never conducted a cross-cultural conversation, the actual usage of emojis in cross-cultural communication is likely to be higher than the survey results suggest.

Still, a vast majority of the participants (92%) believe that emojis can help convey cultural meanings and enhance understanding in cross-cultural communication, and 84% would like to choose emojis considering the cultural background of the recipient during cross-cultural communication in order to better convey their intentions. This shows that even without much experience, participants recognize the potential of emojis to bridge cultural gaps and facilitate cross-cultural communication.

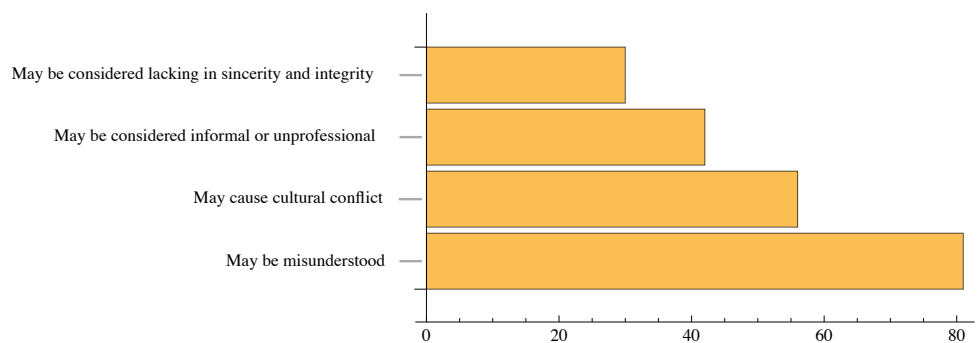

Figure 6: Drawbacks of using emojis in cross-cultural communication

Almost all participants (98%) agree that the meaning of emojis can vary across cultures, which makes common sense as it is similar to gestures having different meanings in different cultural contexts. And this leads to their thoughts on the major drawbacks of using emojis in cross-cultural communication: 76% of the participants believe that emojis may lead to misunderstandings in cross-cultural communication due to differences in interpretation, and 53% believe that it may lead to cultural conflicts.

Another interesting finding is that most of the participants who had never learned about the emojis with cultural characteristics are more likely to show a strong intrest when they are informed about the multi-color options and culture-specific emojis. This suggests that raising awareness about the cultural diversity represented in emojis can enhance users' interest and engagement with these symbols, potentially leading to more inclusive and culturally sensitive communication practices.

# 5 Discussion

## 5.1 Investigation on the evolution process of emojis with cultural characteristics

### 5.1.1 Expansion of multi-color options in emojis

In 1997, Softbank (then known as J-Phone) released the SkyWalker DP-211SW phone, with the world's first known set of emojis (Emojipedia). Then in 1999, Shigetaka Kurita created 176 emojis for NTT DoCoMo's i-mode service (Galloway). Over time, emojis begin to reflect a wider range of cultural characteristics, including different skin tones, genders, foods, and traditional clothing. When it came to 2015, Unicode introduced the skin tone modifiers to indicate the skin tone that should be used for an emoji-style presentation of a pictographic character depicting a person or people, a face, or a hand or body part. Based on the Fitzpatrick skin type classification, five different skin tone options were introduced for emojis.

Table 1: Skin tone modifiers for emojis

| Skin tone modifier | Name | Fitzpatrick scale approximate equivalent |
|---|---|---|
|  | LIGHT SKIN TONE | Types I and II |
|  | MEDIUM LIGHT SKIN TONE | Type III |
|  | MEDIUM SKIN TONE | Type IV |
|  | MEDIUM DARK SKIN TONE | Type V |
|  | DARK SKIN TONE | Type VI |

As emojis continue to evolve, the skin tone options continue to expand. For example, in Emoji 14.0 (released in 2021), Unicode introduced the ability to mix skin tones in a single emoji, allowing for more diverse and inclusive representations (Broni).

These updates reflect a growing awareness of the importance of cultural representation and inclusivity in emoji design, as well as a commitment to providing users with a diverse range of options to express themselves accurately.

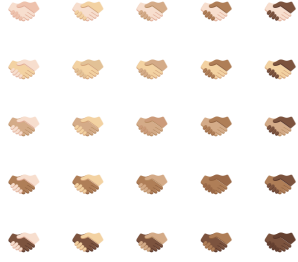

Figure 7: The 5 × 5 combinations of skin tone options in the handshake emoji

### 5.1.2 Expansion of multi-gender options in emojis

Unicode 9.0 (released in 2016) began an effort to improve gender representation in emojis. After the release of Emoji 4.0 (distributed with Unicode 9.0), gender was treated as an explicit property of all humanoid emojis, similar to skin tone. In the first generation of pluralistic emojis, the Unicode Technical Committee partly ignored the recommendations to improve the gender model, resulting in the specification ignoring not only non-binary gender, but many binary gender identities as well. At that time, there were 49 concepts available in all three genders, 26 allowing only two genders, and 8 having only one arbitarily chosen gender (Buff). In Emoji 13.0 (released in 2020), the Unicode Consortium introduced the *inclusive* Zero Width Join (ZWJ) sequence to indicate a genderless emoji symbol, which provides better representation for people who have different gender recognition needs (Daniel).

## 5.2 Analysis: the use of emojis in Eastern and Western cultural backgrounds

In this research project, the survey study and data analysis have revealed the recognition of Chinese college students towards emojis with cultural characteristics, the usage of emojis in cross-cultural communication, and the potential drawbacks of using emojis in cross-cultural communication. The participants also provided insights into the impact of emojis on understanding and interpretation in diverse cultural contexts.

To further explore the differences in emoji usage between Eastern and Western cultural backgrounds, the study by Chandra Guntuku et al. is referred to.

Table 2: Top 15 frequent emojis in East and West (Chandra Guntuku et al.)

| Rank: | 1 | 2 | 3 | 4 | 5 | 6 | 7 | 8 | 9 | 10 | 11 | 12 | 13 | 14 | 15 |
|---|---|---|---|---|---|---|---|---|---|---|---|---|---|---|---|
| **West** |  |  |  |  |  |  |  |  |  |  |  |  |  |  |  |
| | 16.3 | 4.1 | 4.0 | 2.9 | 2.5 | 2.1 | 2.0 | 2.0 | 1.7 | 1.5 | 1.5 | 1.4 | 1.4 | 1.4 | 1.3 |
| **East** |  |  |  |  |  |  |  |  |  |  |  |  |  |  |  |
| | 14.2 | 4.2 | 4.2 | 4.0 | 3.3 | 3.2 | 2.9 | 2.2 | 2.1 | 2.1 | 1.6 | 1.6 | 1.5 | 1.4 | 1.4 |

And here, the following aspects of which influence the use of emojis in Eastern and Western cultural backgrounds are being discussed:

### 5.2.1 The design of emojis influenced by cultural elements

In the East, the prevalence of 🍚 (rice bowl) and 🍜 (ramen) emojis reflects the dietary preferences and cultural values inherent in the region. Conversely, in the West, meat-related emojis (e.g., 🍖, 🍗) are more frequently utilized, indicative of the dietary habits and cultural norms prevalent in those regions. This divergence in emoji usage not only mirrors dietary patterns but also delves into the distinct cultural and social underpinnings of each region.

Within Eastern culture, food is intricately woven into social rituals and familial gatherings, often serving as a symbol of unity and warmth. Consequently, food-related emojis such as 🍲 (hot soup), 🥟 (dumplings), 🍵 (tea), 🥢 (chopsticks), and 🏮 (festive lanterns) are deeply associated with concepts of togetherness and community. The utilization of these emojis in communication not only conveys dietary details but also emotional connections and a sense of belonging.

On the other hand, Western culture places a significant emphasis on meat as a fundamental dietary component, and meat-related emojis are frequently associated with concepts of strength, vitality, and personal indulgence. Emojis depicting 🍖 (grilled meat) and 🥩 (steak) are often viewed as embodiments of individual preference and liberty. In the context of communication, these meat-focused emojis tend to underscore personal tastes and satisfaction, reflecting a culture that values individual expression and choice.

### 5.2.2 Public opinions and governmental policies impacting the use of emojis

In the 1990s in China, video games were widely questioned by the government and public opinion, seen as a negative factor affecting the healthy growth of teenagers. Negative evaluations such as *disrupting social order*, *affecting academic performance*, and *electronic heroin* led to the Chinese public's reservations towards video games. In June 2000, the State Council of China passed an opinion on the special management of video game establishments, explicitly prohibiting the production and sales of gaming devices and their accessories for domestic use (State Council of China). This was followed by restrictions or bans on portable gaming devices, TV gaming consoles, and large gaming arcades. This game console ban was not lifted until 2015, during which time the policy had a significant impact on the popularization and use of gaming culture and related emojis. In the West, due to the free circulation of gaming consoles and the widespread acceptance of gaming culture, 🎮 (gaming controller) is widely used to express "leisure." However, in the East, especially China, due to policy restrictions, these emojis may not have been as common in datasets as early as 2014.

### 5.2.3 Cultural Differences Leading to Interpretation Discrepancies

Under different cultural backgrounds, the same emoji may present completely different interpretations. For example, the emoji 💪, which generally means *strength*, can have vastly different meanings depending on the cultural context. In Canada, people may interpret it as a symbol for showing biceps; in the Arab region, it may be interpreted as an indication of *I have bad body odor*; and for Germans, the emoji may represent the meaning of *Silence is strength*.

Similarly, in American culture, the emoji 🍑 is usually used to express the meaning of *ordinary* or *normal*, but in Japanese social media, the emoji 😇 may imply *bad* or *finished*. In the Middle East, the emoji 👍 is not always used to convey positive and friendly emotions. On the contrary, it may sometimes be seen as an insulting expression.

After in-depth research, the researchers find that there are significant differences between Chinese and Western cultures in terms of values and communication methods.

Western culture generally tends to adopt a direct and clear communication mode, which emphasizes direct expression and clear explanation. In contrast, Eastern culture pays more attention to indirect and implicit expressions, with a focus on the euphemism and suggestion of language. This also fits Hofstede's cultural dimension theory, which divides cultures into high-context and low-context cultures.

In view of the differences in communication styles between high-context culture and low-context culture, relevant studies have provided specific explanations and evidence. A cross-cultural study has collected data from India, Ireland, Thailand, and the United States to delve into the characteristics of communication styles in different cultural contexts (Croucher et al.). The results show that communication styles in high-context countries such as India and Thailand tend to be more conflict-avoidant and submissive, while communication styles in low-context countries such as Ireland and the United States tend to be more intransigent and dominant.

In Eastern cultures, the profound roots of respect culture and high power distance have significant impacts on the use of emojis. Specifically, individuals in Eastern cultures may tend to choose specific types of emojis, such as 🙇, to express respect or apologies, reflecting a respectful and humble attitude towards others. However, in Western cultural contexts, people are generally more inclined to directly use text forms to convey similar concepts rather than relying too much on the assistance of emojis.

In addition, the concept of power distance in Eastern culture is relatively prominent, especially in formal situations such as the workplace. Therefore, in these occasions, the use of emojis is often subject to certain constraints and restrictions to avoid misunderstanding or offense caused by inappropriate use of symbols.

In comparison, the individualistic tendency and equality concept in Western culture provide people with a more relaxed and free environment for the use of emojis. People can flexibly use emojis in various situations, such as using symbols like 👍 to express agreement or support, which fully demonstrates the importance and tolerance of Western culture for personal expression and communication methods.

# 6 Conclusion

Emojis have evolved from a simple set of icons to a complex, culturally rich form of communication that reflects the diversity and inclusivity of the global digital age. This research has highlighted the significant role emojis play in bridging cultural divides, facilitating clearer communication, and fostering inclusivity across varied demographic and cultural landscapes. Through the analysis of emoji usage patterns, cultural representations, and cross-cultural interpretations, it has become evident that emojis are not merely decorative or frivolous symbols; they carry substantial cultural and emotional weight.

The findings suggest that emojis, while enhancing communication, can also lead to misunderstandings if cultural contexts are not considered. This underscores the necessity for continuous evolution in emoji design and selection processes to better accommodate the diverse needs and contexts of global users. The introduction of multi-color options, gender inclusivity, and culturally specific symbols are steps in the right direction, yet the scope for improvement remains vast.

Future research should focus on the development of guidelines for emoji usage in cross-cultural settings to mitigate potential conflicts and misunderstandings. Additionally, exploring the integration of emojis into educational and professional settings could provide insights into their potential to enhance learning and workplace communication.

In conclusion, as digital communication continues to grow, so too should the understanding and implementation of emojis as tools for fostering global connectivity and cultural empathy. Emojis are not just tools of expression but bridges that connect diverse cultures in the expansive digital landscape.

# 8 Appendices

# A Research Proposal

## Project Title

Cross-Cultural Communication in the Digital Age: An Analysis of Cultural Representation and Inclusivity in Emojis

数字时代的跨文化交流：表情符号中的文化表示与包容性分析

## Project Time

Week 7 - Week 16

## Research Background

Emojis, originally created in Japan in the late 1990s, have become a universal language in the digital world. The term "emoji" comes from the Japanese words "e" (絵, meaning "picture") and "moji" (文字, meaning "character"). They have evolved significantly over the years, expanding from a set of 176 icons to over 3,000, reflecting a wide range of emotions, objects, activities, and cultural symbols.

As digital communication has become increasingly global, emojis have incorporated more cultural terms coming from various parts of the world. This includes symbols representing food, festivals, landmarks, and other aspects unique to different cultures. Furthermore, emojis now offer diverse skin tones for people of different ethnicities to better represent themselves, and gender options to promote gender equality and care for the sex minority. There are also options for people with disabilities, reflecting a broader commitment to inclusivity.

However, despite these advancements, the use and interpretation of emojis can still vary significantly across different cultures. This can potentially lead to misunderstandings in cross-cultural communication. Therefore, it is crucial to understand how emojis are used and perceived in different cultural contexts, and how they can facilitate or hinder cross-cultural communication.

## Project Objectives

- To summarize peer-reviewed literature on the evolution of emojis and their role in cross-cultural communication.

- To conduct an analysis on emojis with cultural significance:

  - Investigate the evolution process of emojis with cultural characteristics.

  - Understand the recognition and usage of emojis with cultural characteristics among Chinese university students.

- To conduct an analysis on how emojis are used in cross-cultural communication:

  - Explore the differences in understanding and interpretation of emojis in different cultural contexts.

  - Examine how emojis with cultural characteristics promote the efficiency and inclusiveness of cross-cultural communication.

## Summary Methods

- Conduct a literature review on the evolution of emojis and their role in cross-cultural communication.

- Perform a content analysis of emojis to identify cultural representations.

- Conduct surveys or interviews to gather data on emoji usage and interpretation across different cultures.

## Research Questions

- How have emojis evolved to represent diverse cultures and promote inclusivity?

- How do emojis facilitate cross-cultural communication in the digital age?

- What impact do emojis have on understanding and interpretation in different cultural contexts?

## Project Schedule

| Time | Person | Responsibility |
|------|--------|----------------|
| Week 7-8 | 李柃锋 | Literature review |
| Week 9-10 | 郑翔文 | Content analysis |
| Week 11-12 | Both | Survey, questionnaire distribution and collection |
| Week 13 | 李柃锋 | Data analysis |
| Week 13-14 | 郑翔文 | Discussion and conclusion |
| Week 14 | 李柃锋 | Essay composition, formatting and polishing |
| Week 15 | Both | Final revision, and presentation preparation |
| Week 16 | Both | Presentation |

## Research Prospect

The research aims to contribute to a better understanding of the role of emojis in cross-cultural communication. The findings could have implications for improving digital communication platforms, cross-cultural training programs, and global digital etiquette.

# B Survey Questionnaire

## Emoji（表情符号）在跨文化交流中的使用

**你的年龄段是？[单选题] ***

A. 16 以下

B. 17-25

C. 26-40

D. 40 以上

**您是否使用过 Emoji？（非聊天软件提供的表情符号，是系统/输入法内置的）[单选题] ***

A. 是

B. 否

**您是否知道小黄脸可以长按后选择其它肤色的 Emoji 表情？[单选题] ***

A. 是

B. 否

**您是否注意过 Emoji 中具有文化特征的部分？例如，💂 🧕 🥷 👲 👷 👰 🧟 🦹 👹 🐉 🐍 🐲 🍡 🍘 🥢 🍥 🍲 🍜 🍱 🏔️ 🚉 🏯 🏠 🏡 🚧 🀄 🎎 🎐 🎏 🏮 🔰 📧 等等。[单选题] ***

A. 是

B. 否

**如果您认为还有更多具有文化特征的 Emoji，请在下面打出。[填空题]**

**您是否使用过 Emoji 来进行跨文化交流？[单选题] ***

A. 是

B. 否

**您认为 Emoji 在跨文化交流中有哪些缺点？[多选题] ***

A. 可能会被误解

B. 可能会引起文化冲突

C. 可能会被认为不正式或不专业

D. 可能会被认为缺乏真诚和诚意

**您是否认为 Emoji 在不同文化背景下的含义会有所不同？[单选题] ***

A. 是

B. 否

**您是否会根据对方的文化背景选择使用不同的 Emoji？[单选题] ***

A. 是

B. 否

**您是否认为 Emoji 可以帮助您更好地理解对方的情感和意图？[单选题] ***

A. 是

B. 否

**您是否认为 Emoji 在文化交流中应该受到一定的规范和限制？[单选题] ***

A. 是

B. 否

**您使用具有非中国文化特征的 Emoji 表情的频率如何？ [单选题] ***

A. 很少使用

B. 偶尔使用

C. 经常使用

D. 非常频繁使用

**您使用哪些平台或应用程序进行跨文化交流？ [多选题] ***

A. 微信

B. WhatsApp

C. Facebook

D. Instagram

E. Twitter

F. 其他

G. 不使用

**您是否遇到过因为使用 Emoji 表情而引起的文化误解或冲突？ 如有，请举例说明。[填空题] ***

___________________________________

**计时 30 秒，请尽可能多地在这里打出您认为具有文化特征的所有 Emoji。 [填空题] ***

___________________________________

# 9 Acknowledgements

This paper is formatted with LaTeX and the bibliography was managed with BibTeX; the figures are created with Wolfram Mathematica. The emoji font used is *Twemoji Mozilla*, open-sourced at https://github.com/mozilla/twemoji-colr. The researchers would like to thank the designers and developers of these tools for making the writing and formatting process more efficient.

The researchers claim use of generative artificial intelligence (including large language models such as GPT) in the writing of the paper for summarizing literature and polishing the language. The researchers are responsible for the content of the paper, and are grateful for the assistance provided by these tools.

The researchers would like to thank our research samples for their participation. They have provided valuable insights that have contributed to the findings of this research. The researchers would also like to thank the tutor, Professor Zhu Shanhua, for her guidance and support, and her opening up the world of cross-cultural communication.